\documentclass[12pt]{article}
\usepackage{eurosym}
\usepackage[left=0.7in, right=0.7in, top=0.7in, bottom=0.7in]{geometry}
\usepackage{amssymb,amsmath,amsthm,color,graphicx,graphicx,cite}
\usepackage[caption=false]{subfig}
\usepackage{hyperref}
\usepackage{enumerate}
\usepackage{graphicx}
\usepackage{dsfont}
\usepackage{multirow}
\usepackage{mathrsfs}
\usepackage{times}
\usepackage{bbold}
\usepackage{mathtools}
\usepackage{graphicx}
\usepackage{mathrsfs}
\usepackage{graphicx}
\usepackage{caption}
\usepackage{epstopdf}
\usepackage{epsfig}
\usepackage{eurosym}
\usepackage{booktabs} 
\usepackage{orcidlink}

\begin{document}

	\begin{center}
		\textbf{\Large Detecting Entanglement in High-Spin Quantum Systems via a Stacking Ensemble of Machine Learning Models}
		
		
		\vspace{0.51cm}
	M. Y. Abd-Rabbou $^{a,b\ \orcidlink{0000-0003-3197-4724}}$\footnote{m.elmalky@azhar.edu.eg}, 
	Amr M. Abdallah$^{c\ \orcidlink{0009-0005-4301-9080
	}}$\footnote{12422023414470@pg.cu.edu.eg}, 
	Ahmed A. Zahia \footnote{ahmed.zahia@fsc.bu.edu.eg} $^{d\ \orcidlink{0009-0009-6693-5080}} $\\
	Ashraf A. Gouda$^{b\ \orcidlink{0000-0001-7747-2899}}${\footnote{gouda@azhar.edu.eg}},
	and  Cong‑Feng Qiao $^{a,e\ \orcidlink{0000-0002-9174-7307}}$ \footnote{qiaocf@ucas.ac.cn} \\[0pt]
		\end{center}
		\vspace{0.1cm}
		{\small  $^{a}$School of Physics, University of Chinese Academy of Science, Yuquan Road 19A, Beijing, 100049, China.\\
		$^{b}$Department of Mathematics and Computer Science, Faculty of Science, Al-Azhar University Nasr City, Cairo, 11884, Egypt\\
		$^{c}$ Faculty of Graduate Studies for Statistical Research, Cairo University, Giza 12613, Egypt.\\
		$^{d}$Department of Mathematics, Faculty of Science, Benha University, Benha, Egypt.\\
		$^{e}$ Key Laboratory of Vacuum Physics, University of Chinese Academy of Sciences, Beijing 100049, China.}

	\begin{abstract}
		Reliable detection and quantification of quantum entanglement, particularly in high-spin or many-body systems, present significant computational challenges for traditional methods. This study examines the effectiveness of ensemble machine learning models as a reliable and scalable approach for estimating entanglement, measured by negativity, in quantum systems. We construct an ensemble regressor integrating Neural Networks (NNs), XGBoost (XGB), and Extra Trees (ET), trained on datasets of pure states and mixed Werner states for various spin dimensions. The ensemble model with stacking meta-learner demonstrates robust performance by CatBoost (CB), accurately predicting negativity across different dimensionalities and state types. Crucially, visual analysis of prediction scatter plots reveals that the ensemble model exhibits superior predictive consistency and lower deviation from true entanglement values compared to individual strong learners like NNs, even when aggregate metrics are comparable. This enhanced reliability, attributed to error cancellation and variance reduction inherent in ensembling, underscores the potential of this approach to bypass computational bottlenecks and provide a trustworthy tool for characterizing entanglement in high-dimensional quantum physics. An empirical formula for estimating data requirements based on system dimensionality and desired accuracy is also derived.
		\end{abstract}

\section*{Introduction}
Quantum entanglement, a cornerstone property of quantum physics, underpins advancements in quantum computation and information science. Its conceptual genesis can be traced to seminal debates concerning the completeness of quantum theory, notably Schr\"{o}dinger's articulation in response to the Einstein-Podolsky-Rosen paradox \cite{zhang2024entanglement,bertlmann2023modern}. The quintessential non-local correlations exhibited by entangled states were rigorously scrutinized against local hidden variable theories through the framework of Bell's inequalities, with subsequent experimental verifications, providing compelling empirical evidence refuting local realism \cite{aspect1981experimental,phillips2023experimental}. This affirmation of entanglement's objective physical catalyzed its exploration as an indispensable resource across a spectrum of quantum information processing applications, encompassing quantum computation, secure cryptographic communication, and quantum teleportation \cite{kumar2024secure,singh2023securing}. Consequently, the development of robust methodologies for entanglement detection has become a paramount objective in both experimental and theoretical quantum physics, complemented by the critical challenge of quantifying the degree of entanglement, a task crucial for elucidating a system's intrinsic quantum characteristics and prospective functionalities.

Machine learning models, including NNs, XGB, and ET, have gained prominence as potent analytical instruments for interrogating complex, high-dimensional datasets prevalent in multifaceted scientific and engineering disciplines \cite{nemani2023navigating,wilson2024future,soni2023scalable}. NNs, characterized by their deep architectures and inherent capacity for approximating non-linear functions, demonstrate exceptional efficacy in discerning sophisticated patterns and interdependencies within unstructured data \cite{daubechies2022nonlinear,liu2021adaptive}. XGB, a highly optimized gradient boosting framework, sequentially constructs an ensemble of decision trees, wherein each successive tree is trained to rectify the residual errors of its antecedents, thereby yielding superior predictive accuracy and enhanced robustness against overfitting \cite{thongsuwan2021convxgb,nielsen2016tree}. ET embeds heightened stochasticity within the tree generation process by employing randomized cut-point selection, a strategy that augments model generalization capabilities and improves computational tractability \cite{geurts2006extremely,mastelini2022online}. Transcending the performance limitations of singular models, ensemble learning methodologies assume a pivotal role in augmenting predictive efficacy and operational reliability. Through the aggregation of predictions from multiple base learners, frequently characterized by heterogeneous modeling assumptions, ensemble strategies effectively curtail variance, alleviate overfitting, and encapsulate a more comprehensive spectrum of data patterns \cite{webb2004multistrategy,mienye2022survey}. Methodologies such as bagging, boosting, and stacking epitomize the capacity of ensembles to synergize the distinct strengths of constituent models, thereby achieving superlative performance, particularly in contexts characterized by data complexity, noise, or high dimensionality. Consequently, ensemble-centric approaches have solidified their status as foundational components in contemporary machine learning pipelines, especially for tasks encompassing classification, regression \cite{yu2014improved,zhao2024high}.

The quantification of entanglement, particularly within many-body or high-spin quantum systems, poses a substantial computational impediment to conventional analytical and numerical techniques. This inherent 'computational bottleneck' imposes significant constraints on the exploration of intricate quantum phenomena within high-dimensional regimes, consequently impeding advancements in the comprehension of high-spin physics and its prospective applications. Despite the recognized potential of machine learning in addressing complex problems in quantum physics, a notable gap exists in the development and rigorous evaluation of sophisticated ensemble ML frameworks specifically tailored for the efficient and scalable characterization of entanglement in such high-spin quantum systems. Existing computational methodologies often encounter limitations in terms of accuracy, scalability to higher spin dimensions, or adaptability to the diverse nature of quantum states encountered in these complex regimes, particularly when aiming to bypass traditional numerical bottlenecks without sacrificing predictive power. To address this challenge, leveraging the rapid advancements in machine learning, this study investigates the utility of ensemble machine learning models as an efficient and scalable tool for detecting and estimating entanglement (represented by the negativity measure) in high-spin quantum systems. Our aim is to provide a practical pathway to bypass current computational limitations, thus opening new avenues for research into complex quantum systems. The strongest motivation identifies an important problem, proposes a promising solution (ensemble machine learning), and highlights the broader benefit of this solution.

This paper is structured as follows: Section~\ref{sec:ml_models} details the machine learning models utilized in our ensemble approach, specifically NN, XGB, and ET, and outlines their mathematical formulations. Section~\ref{sec:entanglement_systems} describes the methodology for generating quantum state data for high-spin systems, including pure and Werner states, and introduces the negativity measure of entanglement. Section~\ref{sec:results_discussion} presents and discusses the empirical results, focusing on the performance of the ensemble model in predicting entanglement, its scaling with dataset size and system dimensionality, and comparisons with benchmark states. Finally, Section~\ref{sec:conclusion} summarizes the key findings and delineates potential avenues for future research.

\section{Machine Learning Models for Data Detection} \label{sec:ml_models}

This section outlines the core machine learning models employed in this study: NNs, XGB, and ET. These models are selected for their proven efficacy in discerning complex, non-linear patterns within data. XGB and ET, as ensemble methodologies themselves (comprising multiple decision trees), contribute to the overall predictive framework. The following subsections provide a concise overview of the fundamental mathematical structures and operational principles underlying each approach.

\subsection{Neural Networks}
NNs are computational constructs comprising interconnected nodes, termed neurons, typically organized into an input layer, one or more hidden layers, and an output layer \cite{ellacott2012mathematics}. The operation of an individual neuron is characterized by a weighted sum of its inputs, augmented by a bias term, which is subsequently transformed by a non-linear activation function $\sigma(z)$:
\begin{equation}
	z = \sum_{i=1}^{n} w_i x_i + b, \quad a = \sigma(z)
\end{equation}
where $x_i$ represent input features, $w_i$ are the corresponding weights, and $b$ is the bias.  The network learns by iteratively adjusting its weights and biases to minimize a predefined loss function. For regression tasks, a common choice is the Mean Squared Error (MSE), which quantifies the average squared difference between predicted and true values \cite{sai2009learning}:
\begin{equation}
	\text{MSE} = \frac{1}{N} \sum_{i=1}^{N} (y_i - \hat{y}_i)^2
\end{equation}
where $y_i$ are the true values and $\hat{y}_i$ are the model's predictions for $N$ samples. Optimization is commonly achieved via gradient descent algorithms in conjunction with backpropagation, which computes the gradients of the loss function with respect to each parameter \cite{xu2015multi}:
\begin{equation}
	w_i^{(t+1)} = w_i^{(t)} - \eta \frac{\partial \text{Loss}}{\partial w_i}
\end{equation}
Here, $w_i^{(t)}$ is weight $i$ at iteration $t$, $\eta$ denotes the learning rate, and the last term is the gradient of the  with respect to $w_i$. This iterative learning mechanism enables NNs to model intricate, abstract features from data, conferring robust generalization capabilities across diverse applications.

\subsection{XGBoost}
XGB (Extreme Gradient Boosting) is a highly efficient and scalable implementation of gradient boosting, designed for constructing ensembles of decision trees \cite{yi2024novel}. It operates sequentially, wherein each new tree is trained to correct the residual errors of the preceding ensemble. The objective function $L(\phi)$ minimized at each iteration $t$ combines a loss term $l$ and a regularization term $\Omega(f_k)$ that penalizes model complexity:
\begin{equation}
	L(\phi) = \sum_{i=1}^{n} l(y_i, \hat{y}_i^{(t)}) + \sum_{k=1}^{t} \Omega(f_k)
\end{equation}
The regularization term for an individual tree $f$ is defined as:
\begin{equation}
	\Omega(f) = \gamma T + \frac{1}{2} \lambda \sum_{j=1}^{T} w_j^2
\end{equation}
where $T$ is the number of leaves, $w_j$ are the scores on each leaf, and $\gamma, \lambda$ are regularization parameters. XGB employs sophisticated techniques, including second-order Taylor expansion for loss approximation and efficient handling of sparse data, contributing to its superior performance and speed in numerous applications.

\subsection{Extra Trees}
The ET algorithm is an ensemble learning technique that aggregates predictions from multiple decision trees characterized by a high degree of randomization \cite{geurts2006extremely,mastelini2022online}. Unlike traditional tree-based ensembles, ET introduces randomness in two primary ways: by randomly selecting a subset of features at each node and by choosing split points for these features completely at random from their range within the current node's data. Trees are constructed using the entire training dataset (i.e., without bootstrapping). For regression tasks, the optimal split is typically selected based on its ability to maximize variance reduction. The variance reduction $\Delta \text{Var} (t)$ at a node $t$ when split into left ($t_L$) and right ($t_R$) child nodes is given by:
\begin{equation}
	\Delta \text{Var}(t) = \text{Var}(t) - \left( \frac{n_L}{n_t} \text{Var}(t_L) + \frac{n_R}{n_t} \text{Var}(t_R) \right)
\end{equation}
where $\text{Var}(t)$ is the variance of the target variable at node $t$, and $n_t, n_L, n_R$ are the number of samples at node $t$ and its respective children. The high degree of randomization in ET often leads to reduced variance in the final ensemble prediction and computational efficiency, particularly for high-dimensional datasets.

\subsection{CatBoost}
CatBoost (Categorical Boosting) is a gradient boosting algorithm developed by Yandex, specifically optimized to handle categorical features with minimal preprocessing \cite{hancock2020catboost,ibrahim2020comparison}. Unlike traditional gradient boosting methods, CatBoost uses \textit{ordered boosting} and \textit{target statistics} to reduce overfitting and handle categorical variables natively. The fundamental goal of CatBoost is to minimize a loss function \( L(y, \hat{y}) \), where \( y \) is the true label and \( \hat{y} \) is the predicted output. It constructs an ensemble of decision trees \( F(x) = \sum_{m=1}^M f_m(x) \), where each tree \( f_m(x) \) is trained to correct the error of the previous ensemble:

\[
\hat{y}^{(m)} = \hat{y}^{(m-1)} + \eta \cdot f_m(x)
\]

where \( \eta \) is the learning rate, and \( \hat{y}^{(m)} \) is the prediction at iteration \( m \). One of CatBoost's main strengths is its approach to categorical features. Instead of one-hot encoding, it employs \textit{target statistics} using random permutations to avoid target leakage. For a categorical feature \( x_c \), the transformation is defined as:

\[
\text{TS}(x_c) = \frac{\sum_{i=1}^{t-1} \mathbb{I}(x_i = x_c) y_i + a \cdot P}{\sum_{i=1}^{t-1} \mathbb{I}(x_i = x_c) + a}
\]

where \( \mathbb{I}(\cdot) \) is the indicator function, \( a \) is a regularization parameter, \( P \) is the prior mean (typically the global mean of the target), and \( t \) is the index in a random permutation of the data. This efficient handling of categorical features, combined with robust boosting strategies, makes CatBoost highly accurate and suitable for a wide range of machine learning tasks.

\subsection{Ensemble Framework}
Ensemble stacking, also known as stacked generalization, is a machine learning technique that combines multiple base models to improve predictive performance. The fundamental idea is to train a set of diverse base learners \( f_1(x), f_2(x), \ldots, f_n(x) \) on the training data, and then use their predictions as inputs to a higher-level model called the meta-learner. Instead of relying on a simple average or majority vote, stacking allows the meta-learner to learn the optimal combination of base model outputs. Formally, the final prediction is given by:

\[
\hat{y} = g\left(f_1(x), f_2(x), \ldots, f_n(x)\right)
\]

where \( f_i(x) \) is the prediction of the \( i^{\text{th}} \) base model for input \( x \), and \( g(\cdot) \) represents the meta-learner. In this setup, the base models consist of a NN, ET, and XGB. These models are chosen for their diverse learning mechanisms, allowing them to capture different aspects of the data. The meta-learner, tasked with intelligently learning how to combine their predictions, is CatBoost. To prevent data leakage and reduce overfitting, the meta-learner is not trained on the raw data. Instead, it is trained on the out-of-fold predictions generated by the base learners during a cross-validation procedure. This ensures the meta-learner learns a robust mapping from the base model predictions to the true labels, allowing the final ensemble to capture complex relationships between model outputs and true labels, resulting in improved accuracy and generalization.

\begin{figure}[h!]
	\centering
	\includegraphics[width=1\textwidth]{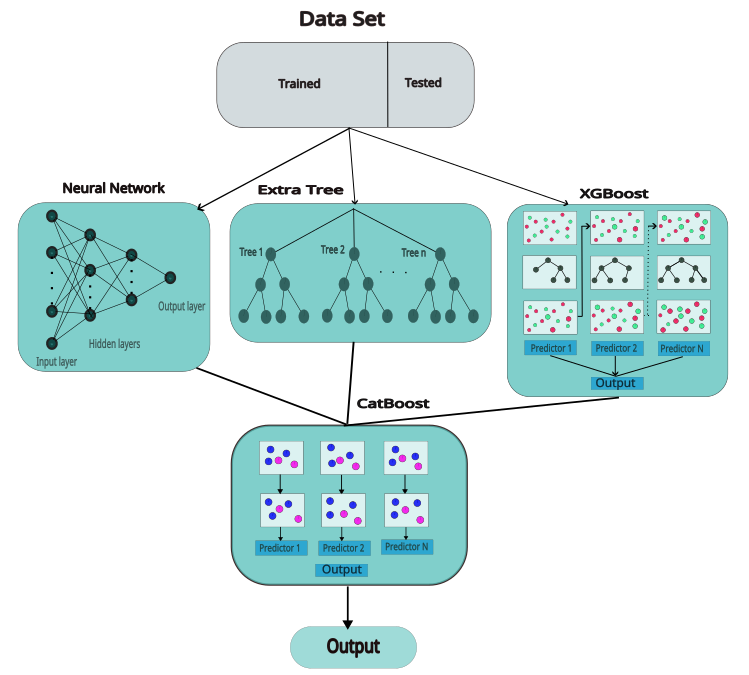}
	\caption{The dataset is split into $80\%$ for training and $20\%$ for testing. An ensemble model using stacking is constructed with three base learners( Neural Networks, Extra Trees, and XGBoost) each trained independently. Their predictions are used as input features for the meta-learner (CatBoost), which learns to optimally combine these predictions to produce the final, more accurate output}
	\label{fig1}
\end{figure}

\section{Entanglement for High-Spin Quantum Systems} \label{sec:entanglement_systems}
 Among the various quantifiers of entanglement, negativity stands out as a particularly potent and computationally tractable measure, especially for bipartite quantum systems \cite{vidal2002computable}. Negativity quantifies the extent to which the partial transpose of a density matrix fails to remain positive semidefinite, thereby capturing essential entanglement characteristics of a quantum state. For a bipartite density matrix $\rho_{AB}$ describing a system composed of subsystems A and B, the negativity $\mathcal{N}(\rho_{AB})$ is formally defined as \cite{calabrese2012entanglement}
\begin{equation}
	\mathcal{N}(\rho_{AB}) = \frac{\|\rho_{AB}^{T_B}\|_1 - 1}{2}
\end{equation}
where $\rho_{AB}^{T_B}$ denotes the partial transpose of $\rho_{AB}$ with respect to subsystem B, and $\|\cdot\|_1$ signifies the trace norm (i.e., the sum of the singular values of the matrix). A state $\rho_{AB}$ is classified as separable if $\mathcal{N}(\rho_{AB}) = 0$. For normalized negativity (ranging from 0 to 1, depending on the system dimension and normalization convention), a value of 1 can indicate maximal entanglement, while $0 < \mathcal{N}(\rho_{AB}) < 1$ signifies partial entanglement.

Negativity serves as a robust entanglement monotone, adept at detecting non-classical correlations, particularly in mixed quantum states where other measures, such as concurrence, may be less effective or more challenging to compute. A significant advantage of negativity is its applicability to systems of arbitrary Hilbert space dimensions, offering enhanced versatility and robustness for characterizing entanglement in high-dimensional quantum systems \cite{adesso2016measures}. Furthermore, negativity has demonstrated efficacy in revealing entanglement properties that might be obscured under local operations and classical communication. Its sensitivity to quantum correlations and resilience against decoherence and environmental noise underscore its utility in the study and advancement of complex quantum systems, often rendering it superior to concurrence for both detection and quantification tasks in such regimes.

\subsection{Generating Quantum State Data and Training the Ensemble Model}
This subsection delineates the methodology for constructing the quantum state datasets essential for training and rigorously evaluating the proposed machine learning models. The primary focus is on bipartite quantum systems characterized by high-dimensional Hilbert spaces, reflecting the challenges posed by high-spin configurations.

The foundational representation for a pure bipartite quantum state $|\psi\rangle$, involving two subsystems with total angular momenta $J_1$ and $J_2$ respectively, is expressed as:
\begin{equation}
	|\psi\rangle = \sum_{m=-J_1}^{J_1} \sum_{n=-J_2}^{J_2} C_{mn} |m, n; J_1, J_2\rangle
\end{equation}
In our primary dataset generation protocol, the amplitudes $C_{mn}$ are initially sampled as real numbers from a standard Gaussian distribution with zero mean, $\mu = 0$, and unit standard deviation, $\sigma = 1$. Subsequently, the state vector $|\psi\rangle$ is normalized. This procedure is a common method for approximating a uniform (Haar) distribution over the space of pure quantum states with real coefficients, ensuring a diverse and representative sampling. The input features for the machine learning models are constructed directly from these real amplitudes. For each state, the set of $C_{mn}$ coefficients is concatenated to form a single real-valued feature vector. The dimensionality of this vector is $(2J_1+1) \times (2J_2+1)$, corresponding to the total number of amplitudes. This method preserves the crucial sign information of the amplitudes, which would be lost if only their magnitudes were used, and provides a complete description of the real-valued state vector. The corresponding target variable for supervised learning is the negativity $N(\rho_{AB})$ calculated for the density matrix $\rho_{AB} = |\psi\rangle\langle\psi|$.

A dataset comprising $S$ random quantum states was meticulously constructed. To address potential class imbalance and ensure robust learning, particularly for distinguishing entangled from separable states, the dataset was structured such that approximately 10\% of the states were generated to be demonstrably entangled (by ensuring a sufficient number of non-zero, appropriately distributed $C_{mn}$ amplitudes leading to non-zero negativity). The remaining 90\% of states were predominantly separable or exhibited very low entanglement. This latter category was generated by systematically varying the sparsity of the $C_{mn}$ vectors: for different groups of states, a controlled number of $C_{mn}$ amplitudes (with randomly chosen indices and values) were set to be non-zero, ranging from a single non-zero amplitude (guaranteeing separability) up to a threshold that typically maintains separability or minimal entanglement. This strategy ensures a rich spectrum of non-entangled and weakly entangled states, providing challenging negative examples for the learning algorithms.

To stringently evaluate the model's predictive capabilities on well-characterized quantum states, two specific families were incorporated:
\begin{enumerate}
	\item A class of tunable pure entangled states $|\zeta(\theta)\rangle$, defined for $J_1=J_2=J$ as:
	\begin{equation}
		|\zeta(\theta)\rangle = \cos(\theta) |-J, -J\rangle + \sin(\theta) |J, J\rangle
	\end{equation}
	where the parameter $\theta \in [0, \pi/2]$ allows continuous variation of entanglement from separable (at $\theta = 0, \pi/2$) to maximally entangled (at $\theta = \pi/4$). This serves as a critical benchmark for pure state entanglement prediction.
	\item Werner states, a canonical example of mixed states, valuable for assessing model robustness to statistical mixtures and noise:
	\begin{equation}
		\rho_W = \alpha |\phi\rangle\langle\phi| + \frac{1-\alpha}{D^2} I
	\end{equation}
	where $\alpha \in [0, 1]$, $D = (2J+1)$ is the dimension of each subsystem (for $J_1=J_2=J$), $I$ is the identity operator on the composite Hilbert space $\mathcal{H}_1 \otimes \mathcal{H}_2$, and $|\phi\rangle$ represents a maximally entangled pure state, typically the Bell-like state:
	\begin{equation}
		|\phi\rangle = \frac{1}{\sqrt{D}} \sum_{k=-J}^{J} |k, k; J, J\rangle
	\end{equation}
	The parameter $\alpha$ interpolates between a pure maximally entangled state ($\alpha=1$) and a maximally mixed, separable state ($\alpha=0$).
\end{enumerate}

Prior to training, all input features underwent standardization (scaling to zero mean and unit variance) to ensure numerical stability and improve convergence during the optimization process. This preprocessing step was applied to the vector of real-valued $C_{mn}$ amplitudes for the pure state and the Werner state datasets. The machine learning models were configured as follows:
\begin{itemize}
	\item \textbf{Neural Network}: A feedforward, fully connected architecture implemented using Keras' Sequential API. It featured an input layer conforming to the feature vector dimension, followed by three hidden layers with 128, 64, and 32 neurons, respectively. All hidden layers utilized the rectified linear unit activation function for introducing non-linearity. The output layer comprised a single neuron with a linear activation function, appropriate for the regression task of predicting negativity.
	\item \textbf{XGB}: Configured for regression using the `reg:squarederror` objective function, with `n\_estimators=300` (number of boosting rounds), `learning\_rate=0.05` (step size shrinkage), `max\_depth=10` (maximum tree depth), and `random\_state=42` to ensure reproducibility of results.
	\item \textbf{ET}: Implemented with `n\_estimators=300` (number of trees in the forest), `max\_depth=15` (maximum depth of individual trees), `random\_state=42` for consistency, and `n\_jobs=-1` to leverage all available CPU cores for parallel computation, enhancing training efficiency.
	\item \textbf{CB}: The model is set to train with up to 1000 estimators, using early stopping after 50 rounds without improvement to prevent overfitting and speed up training. A fixed `random\_state` of 42 ensures reproducibility, while `verbose=100` provides progress updates every 100 iterations, making it easier to monitor model training. This setup balances accuracy, efficiency, and transparency during the training process.
	
\end{itemize}
The comprehensive dataset was partitioned using a standard 80\% for training the models and the remaining 20\% for rigorous, unseen testing and performance evaluation, as conceptually illustrated in Figure \ref{fig1}. The overarching goal of the training process was to minimize the discrepancy between the model's predicted negativity and the true, calculated negativity for the quantum states in the training set.

\section{Results and Discussion} \label{sec:results_discussion}
This section presents a comprehensive analysis of the ensemble machine learning model's performance in predicting entanglement, specifically the negativity, for multi-spin quantum systems. The ensemble architecture is a \textbf{stacking model} that amalgamates predictions from three diverse base learners: a NN, ET, and XGB. The predictions from these base models are then used as input features for a higher-level \textbf{meta-learner, CB}, which produces the final output. This stacking strategy is designed to capitalize on the distinct strengths inherent to each constituent model. The meta-learner's task is to learn the optimal way to combine the base predictions, effectively weighting their contributions based on their performance across the problem space, rather than relying on a simple average or manual weighting. The tree-based methods (ET and XGB) complement the NN by enhancing robustness and providing alternative perspectives on feature importance, while the NN excels at capturing complex, non-linear interdependencies within the data. The rationale for this ensemble construction is further discussed in Appendix~\ref{sec:appendix_model_comparison}. This ensemble approach aims to strike a balance between the representational flexibility of deep learning and the interpretative stability often associated with ensemble tree models, thereby fostering improved predictive accuracy for the regression tasks under investigation.  

\subsection{Performance Scaling with Dataset Size and Derivation of an Empirical Formula}
\begin{figure}[!h]
	\centering
	\includegraphics[width=0.95\textwidth,height=390pt]{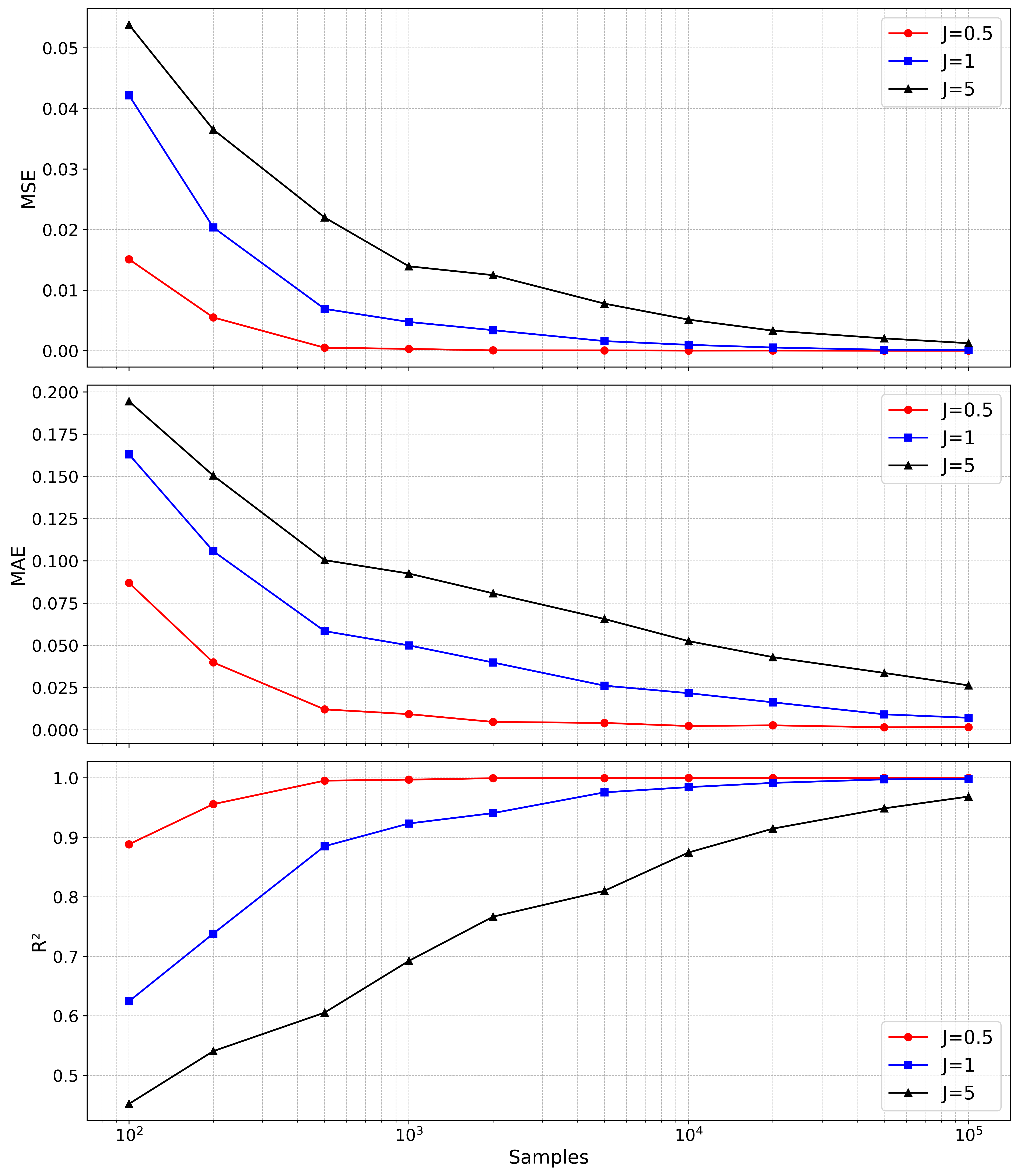}
	\caption{Pure state metrics with respect to the number of random states in the case of pure states}
	\label{sample_pure}
\end{figure}

A critical aspect of evaluating any machine learning model is understanding its performance scalability with respect to the volume of training data, particularly when addressing problems like entanglement quantification that suffer from the "computational bottleneck" inherent in high-dimensional quantum systems. Figure \ref{sample_pure} provides a detailed examination of this relationship for the ensemble model when applied to pure quantum states, across varying spin dimensions characterized by $J = 0.5, 1,$ and $5$. The analysis focuses on three standard performance metrics: MSE, Mean Absolute Error (MAE), and the coefficient of determination ($R^2$).

As anticipated from established machine learning principles, the results consistently demonstrate that increasing the number of training samples leads to an enhancement in model performance across all considered spin values. This is manifested by a monotonic decrease in both MSE (Figure \ref{sample_pure}, top panel) and MAE (Figure \ref{sample_pure}, middle panel), coupled with a corresponding monotonic increase in the $R^2$ score (Figure \ref{sample_pure}, bottom panel), as the sample size (plotted on a logarithmic x-axis) expands. This trend underscores the model's ability to learn underlying patterns more effectively and generalize better when provided with more extensive training data.

A salient observation is the pronounced dependence of convergence speed and ultimate performance on the dimensionality of the quantum system, represented by the spin quantum number $J$. For lower-dimensional systems (e.g., $J = 0.5$, indicated by red circles), the model exhibits rapid performance improvements. Error metrics decrease sharply, and the $R^2$ value swiftly approaches unity, indicating near-optimal performance achieved with a relatively modest number of samples (on the order of $10^3$). The $J = 1$ system (blue squares) follows a similar trajectory but necessitates a moderately larger dataset to attain comparable levels of accuracy.

In stark contrast, the high-dimensional $J = 5$ system (black triangles) demonstrates significantly slower convergence. The MSE and MAE decrease more gradually, and the $R^2$ score ascends at a much-reduced rate. This signifies a substantially greater data requirement to achieve high-fidelity entanglement predictions for higher-spin systems. This behavior is intrinsically linked to the exponential growth of the Hilbert space dimension with $J$, leading to a vastly more complex state space and entanglement landscape that the model must learn to navigate. Even at the maximum depicted sample size of $10^5$, the $R^2$ for $J = 5$ is still visibly improving and has not yet reached an asymptotic performance level, unlike the curves for $J = 0.5$ and $J = 1$, which exhibit saturation or diminishing returns at smaller sample sizes. This highlights the "curse of dimensionality" impacting the data efficiency of the learning process.

\begin{figure}[h!]
	\centering
	\includegraphics[width=0.95\textwidth,height=390pt]{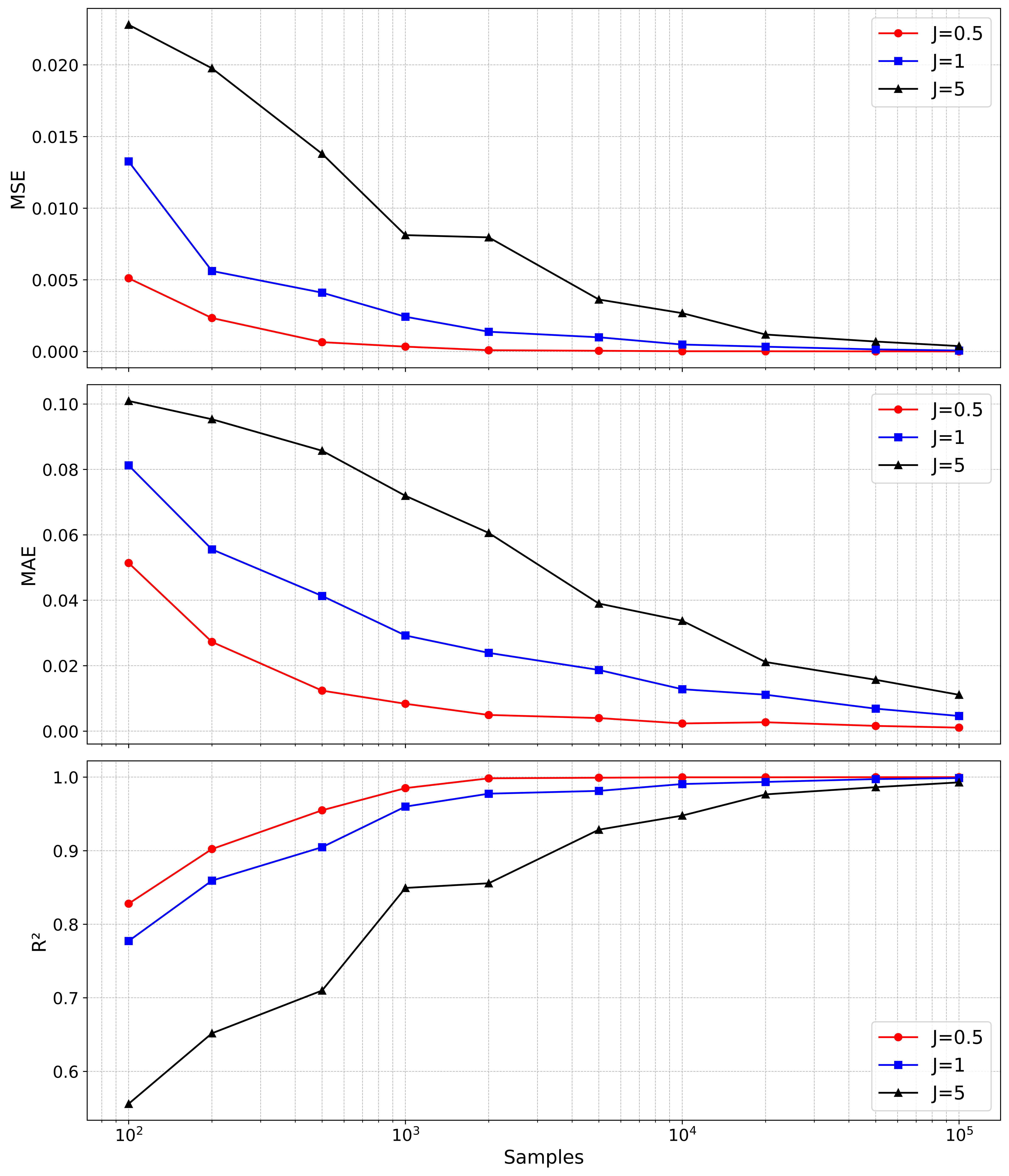}
	\caption{Werner states metrics with respect to the number of random states in the case of Werner states.}
	\label{sample_werner}
\end{figure}

Figure \ref{sample_werner} extends this scalability analysis to Werner states, which serve as a proxy for more realistic mixed quantum systems incorporating noise and statistical mixtures. The plots again illustrate MSE, MAE, and $R^2$ as functions of training sample size (logarithmic scale) for spin dimensions $J = 0.5, 1,$ and $5$. Consistent with observations for pure states, increasing the sample size invariably leads to improved predictive accuracy across all dimensions. Both MSE and MAE exhibit a clear downward trend, while $R^2$ correspondingly increases.

The rate of performance improvement for Werner states is also strongly contingent on the system's dimensionality ($J$). The $J = 0.5$ system (red circles) demonstrates the most rapid convergence. The $J = 1$ system (blue squares) requires a larger dataset (approximately $10^4$ samples) to reach comparable saturation levels. The high-dimensional $J=5$ case (black triangles) presents a considerably greater challenge, with its convergence being markedly slower. The increased complexity arising from mixedness, compounded by higher dimensionality, means that even at $10^5$ samples, the $R^2$ for $J = 5$ (while high at nearly 0.98) appears to be still ascending, suggesting that optimal characterization of entanglement in such high-dimensional mixed states might demand even larger datasets.

Overall, Figures \ref{sample_pure} and \ref{sample_werner} confirm the ensemble model's robustness in handling both pure and mixed quantum states. They underscore that while the ML approach shifts the computational burden from direct calculation on individual states to data generation and model training, the training data requirement itself scales significantly with system complexity. This empirically justifies the strategic use of progressively larger datasets for higher $J$ values.

Leveraging these observed performance trends, we sought to derive a phenomenological scaling law to estimate the number of training samples ($S$) required. A linear regression model was applied to the aggregated data from Figures \ref{sample_pure} and \ref{sample_werner}, using performance metrics (MSE, MAE, $R^2$) and the spin quantum number $J$ as input features, with $\log_{10}(S)$ as the target variable. This analysis yielded the following empirical relationship:
\begin{equation}
	\log_{10}(S) \approx 2.8 + (0.502)\ J - (3.042) \ \text{MSE} - (8.012) \ \text{MAE} + (1.012)\  R^2
	\label{eq:empirical_formula}
\end{equation}
This heuristic formula provides a quantitative guideline for estimating data requirements. The positive coefficient for $J$ (+0.502) quantitatively confirms that the necessary sample size $S$ grows exponentially with the spin quantum number, as expected from the "curse of dimensionality." It is insightful to consider this scaling in the context of the Hilbert space dimension, $D = (2J+1)^2$. The logarithm of the number of samples, $\log_{10}(S)$, appears to scale roughly linearly with $J$. This empirical finding suggests that the complexity of the learning task, while substantial, may not scale directly with the full state space dimension (which would be closer to $\log(D) \approx 2\log(2J+1)$ for large $J$), but rather with a simpler proxy for its complexity. The signs of the other coefficients are also intuitive: achieving lower errors (MSE, MAE) requires more samples, hence the \textbf{negative coefficients} for these terms, while achieving a higher coefficient of determination ($R^2$) also requires more samples, as indicated by its positive coefficient. While this formula is an empirical approximation specific to our ensemble architecture and dataset characteristics and should be applied cautiously outside the investigated parameter range, it offers a practical tool for initial resource estimation in similar studies. The dataset sizes employed elsewhere in this study (e.g., $10^4$ for $J=0.5$, $2 \times 10^4$ for $J=1$, and $10^5$ for $J=5$) were guided by these scaling observations, aiming for a balance between predictive accuracy and computational tractability.

\subsection{Performance Evaluation for Pure Quantum States}

\begin{figure}[h!]
	\centering
	\includegraphics[width=0.32\textwidth]{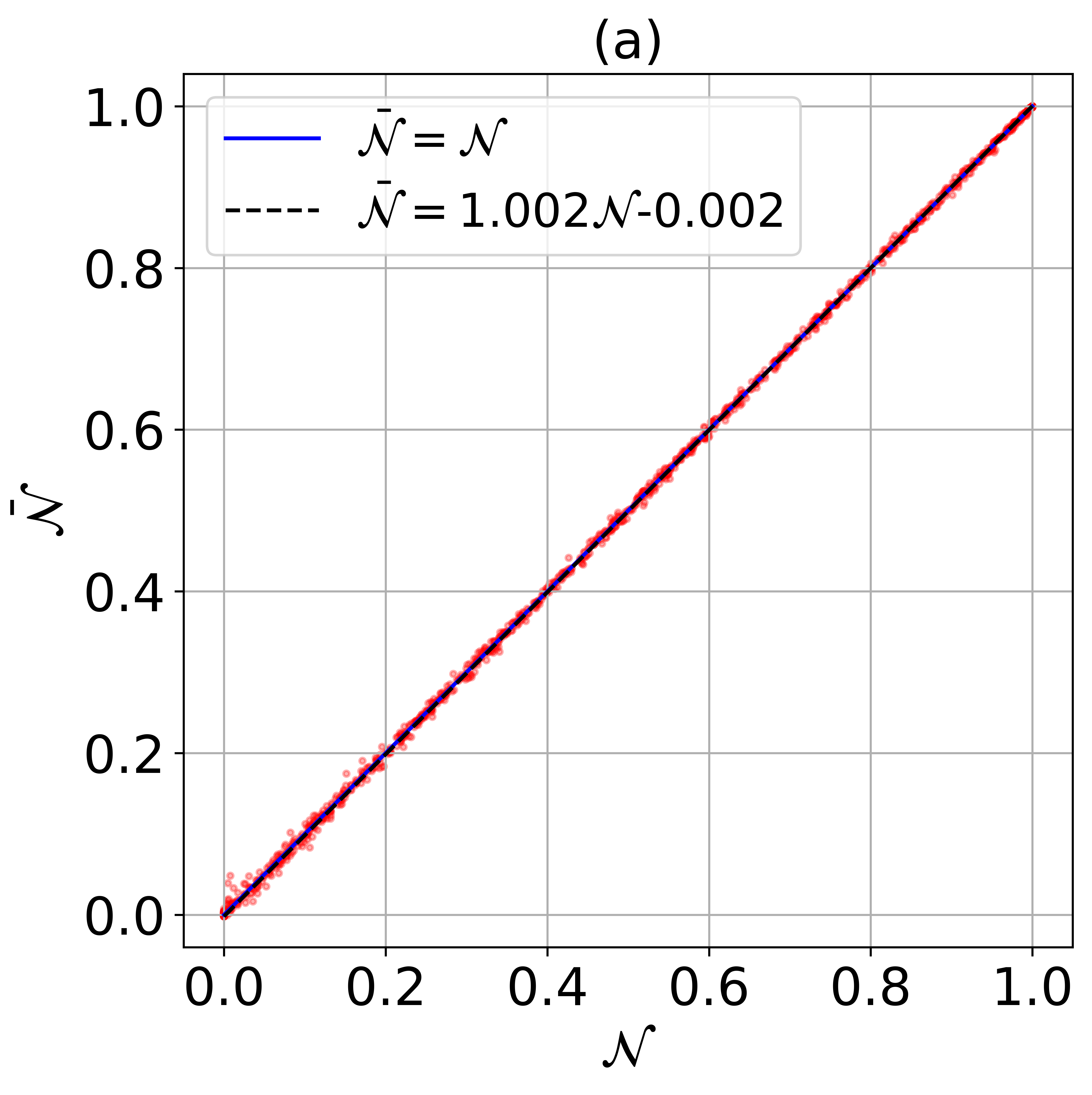}
	\includegraphics[width=0.32\textwidth]{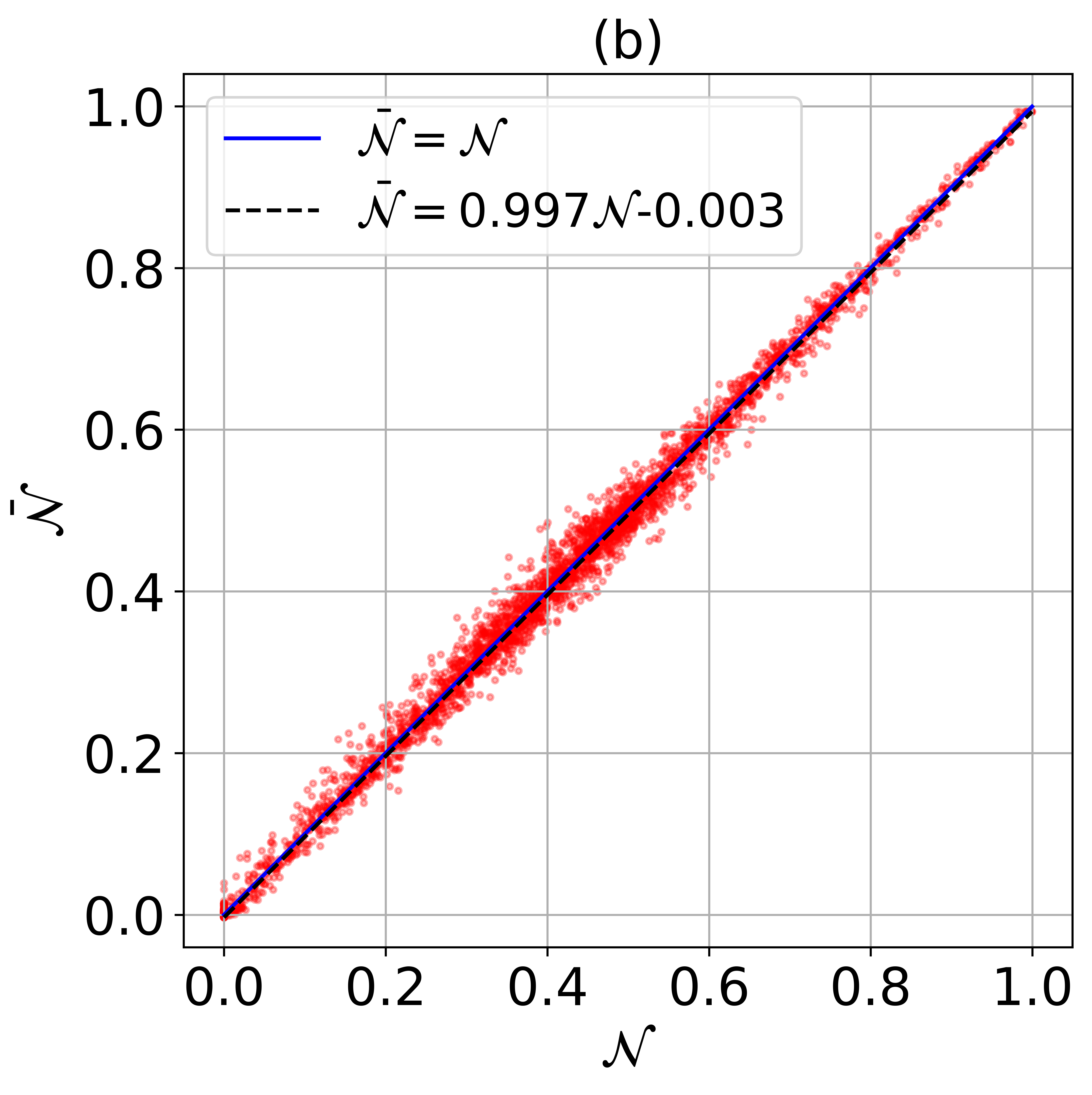}
	\includegraphics[width=0.32\textwidth]{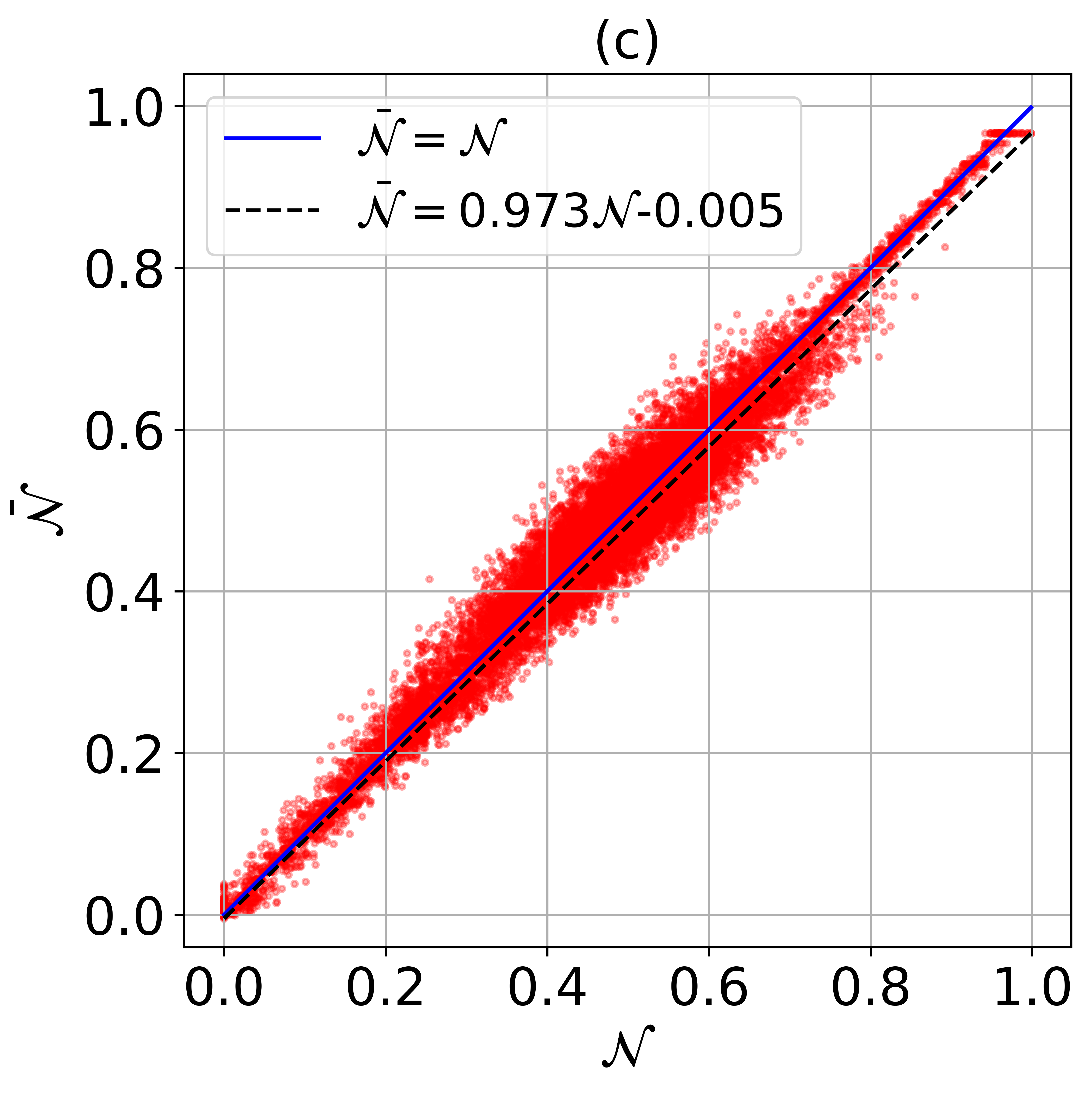} \\
	\includegraphics[width=0.32\textwidth]{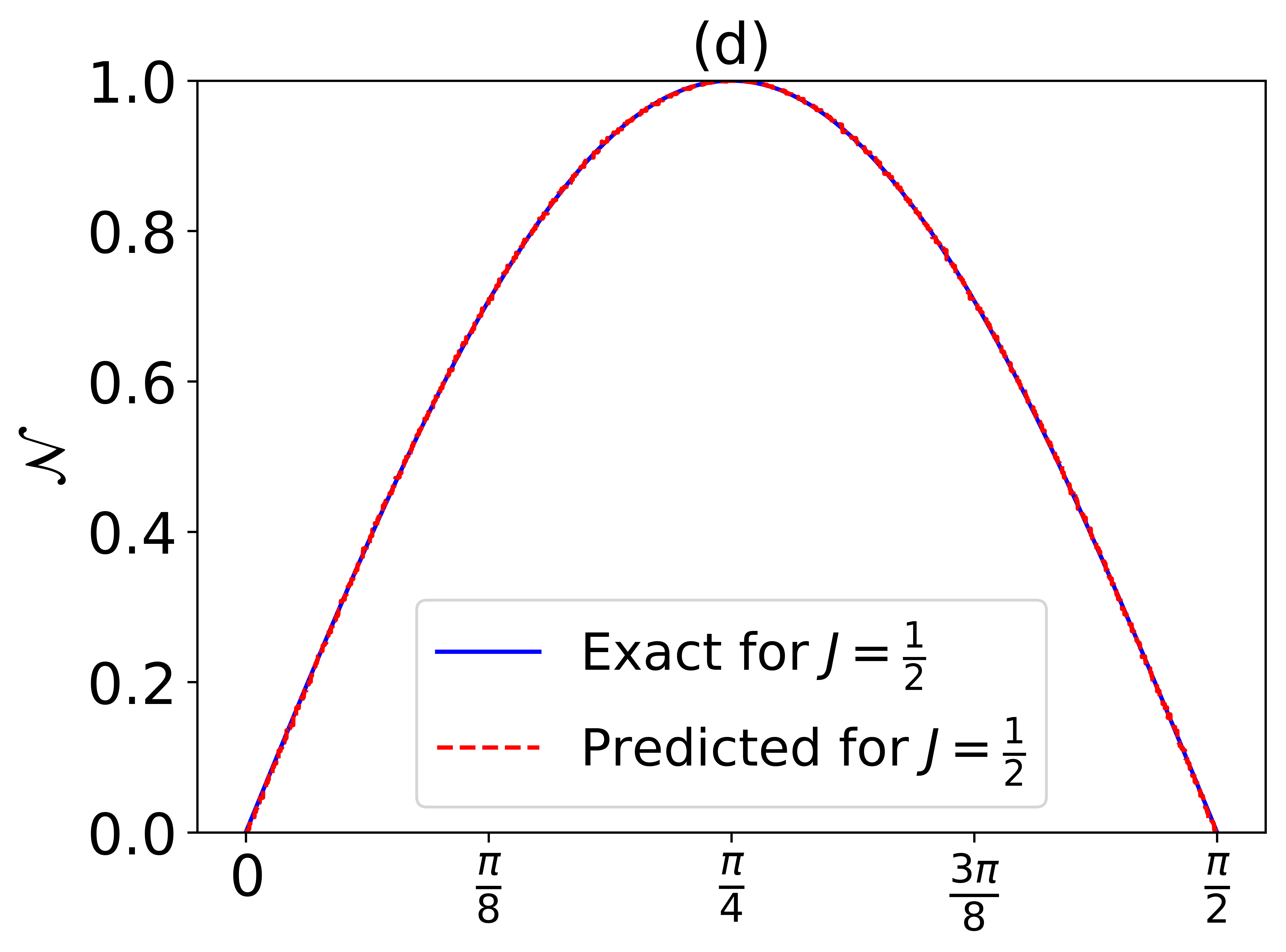}
	\includegraphics[width=0.32\textwidth]{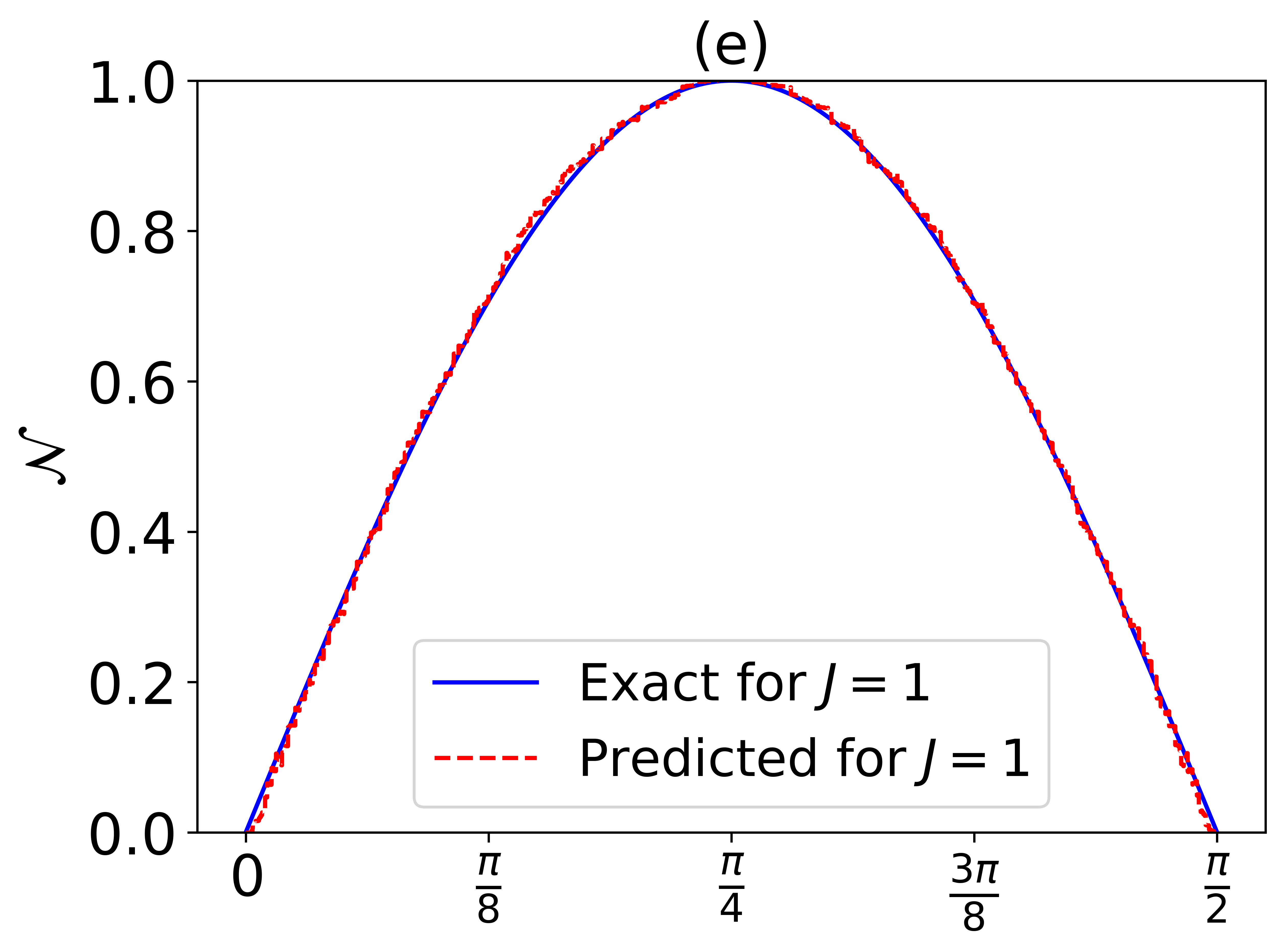}
	\includegraphics[width=0.32\textwidth]{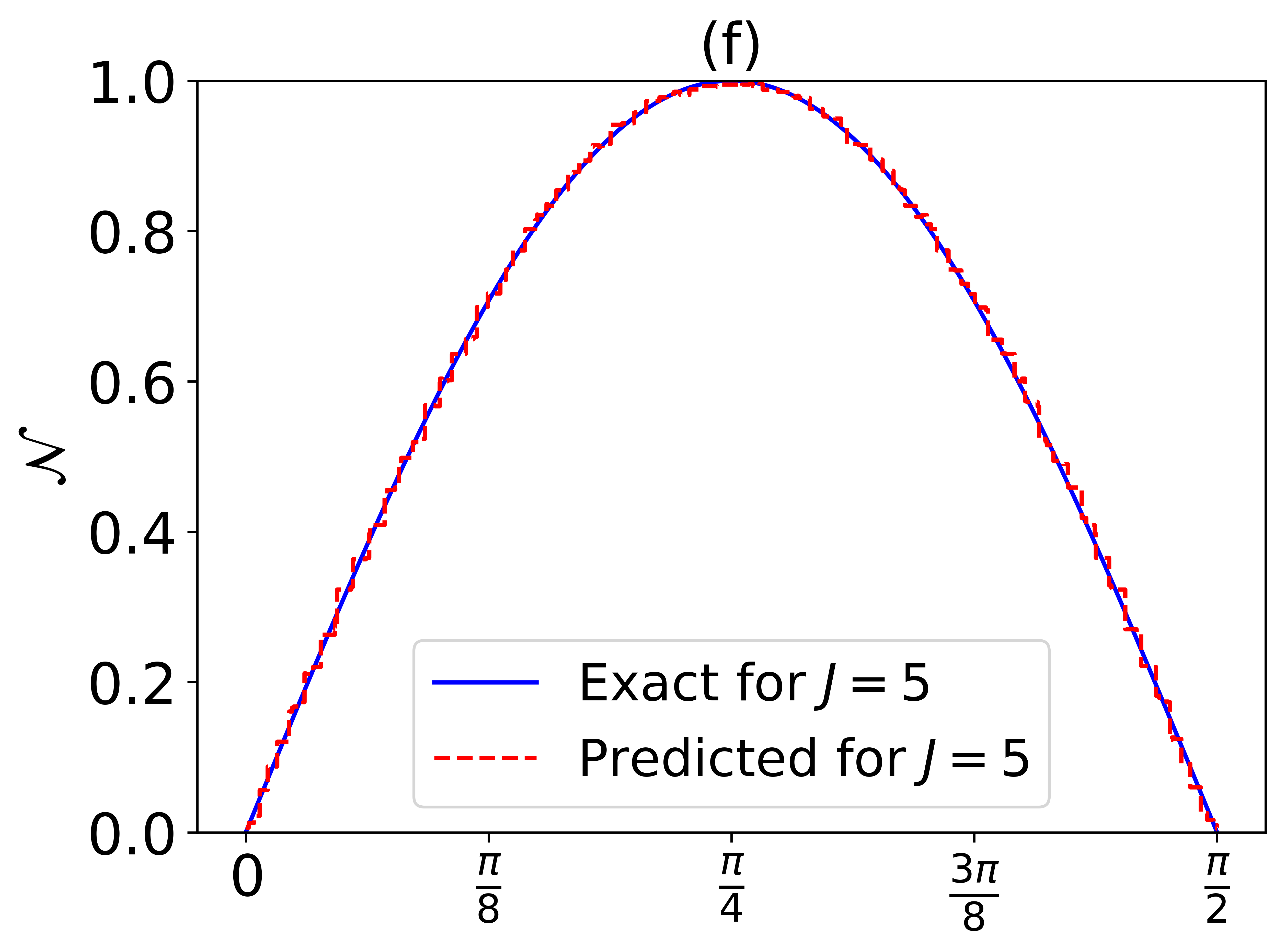}
	\caption{Upper plates: Comparison of actual and predicted values for pure state  (a) \( J = \frac{1}{2} \), (b) \( J = 1 \), (c) \( J =5\). Lower plates: Testing the ensemble model for pure state  (d) \( J = \frac{1}{2} \), (e) \( J = 1 \), (f) \( J =5\).  }
	\label{f33}
\end{figure}

This subsection presents a detailed evaluation of the ensemble machine learning model's predictive accuracy for the negativity ($\mathcal{N}$) of pure quantum states across different spin dimensions ($J=0.5, 1,$ and $5$). Pure states, by their nature, lack statistical mixing, potentially allowing for a more direct mapping from their state vector components (the input features $C_{mn}$) to their entanglement properties, which the model aims to learn. The analysis encompasses both a general assessment of randomly generated pure states and a focused examination on a specific, parametrically tunable family of pure states, the latter providing a more stringent test of the model's ability to capture underlying physical relationships.

Figure \ref{f33} (upper panels) provides a direct comparison between the actual, calculated negativity values and those predicted by the ensemble model for randomly generated pure states. For $J=0.5$ (qubit-qubit systems, Figure \ref{f33}a), the model demonstrates exceptional predictive fidelity. The scatter plot of predicted negativity ($\tilde{\mathcal{N}}$) versus true negativity ($\mathcal{N}$) shows data points tightly clustered along the ideal $\tilde{\mathcal{N}} = \mathcal{N}$ line. This high degree of correlation is quantitatively substantiated by a coefficient of determination ($R^2$) of $0.9999$. A linear regression fit to these predictions yields $\tilde{\mathcal{N}} \approx 1.002\mathcal{N} - 0.002$, indicating a near-perfect linear relationship with negligible systematic deviation. This performance underscores the model's profound capability to accurately characterize entanglement in low-dimensional pure states. For $J=1$ (qutrit-qutrit systems, Figure \ref{f33}b), the ensemble model continues to exhibit robust predictive performance. While a marginal increase in prediction variance (scatter) is observable compared to the $J=0.5$ case—attributable to the larger Hilbert space and correspondingly more complex feature space—the data points remain predominantly aligned with the $\tilde{\mathcal{N}} = \mathcal{N}$ diagonal. The $R^2$ value remains impressively high at $0.9962$. The linear regression fit, $\tilde{\mathcal{N}} \approx 0.997\mathcal{N} - 0.003$, shows a slight increase in slope and intercept deviation, reflecting the increased complexity associated with the higher-dimensional qutrit system. Nevertheless, the model's accuracy is substantial. As the system dimensionality escalates significantly, $J=5$, the challenge for the model intensifies. Figure \ref{f33}c reveals a more pronounced scatter of predicted values around the ideal line. This increased variance is an expected consequence of navigating a vastly larger feature space where entanglement can manifest in more intricate ways, demanding more data for precise characterization. Despite this, the model achieves a strong $R^2$ of $0.9717$. The linear regression fit, $\tilde{\mathcal{N}} \approx 0.973\mathcal{N} - 0.005$, remarkably indicates minimal systematic bias. This demonstrates the model's commendable ability to generalize to high-spin pure states, effectively learning the salient features indicative of entanglement even in this complex regime.

To further probe the model's capacity to learn the functional dependence of entanglement on state parameters—a more rigorous test than average performance on random states—its performance was evaluated on the parametrically tunable pure state $|\zeta(\theta)\rangle = \cos(\theta)|-J,-J\rangle + \sin(\theta)|J,J\rangle$, where $\theta$ continuously varies the entanglement (see Equation 9). The results, presented in Figure \ref{f33} (lower panels), compare the model's predicted negativity against the exact theoretical negativity as a function of $\theta$.
For $J=0.5$ (Figure \ref{f33}d), the model's predictions (dashed red line) achieve a near-perfect overlay with the exact theoretical curve (solid blue line) across the entire parameter range of $\theta$. This demonstrates that the ensemble model has effectively learned the precise functional relationship governing entanglement in this qubit system, including accurate predictions at the separable and maximally entangled points. For $J=1$ (Figure \ref{f33} e), the predicted negativity for the tunable state $|\zeta(\theta)\rangle$ exhibits remarkable fidelity, closely tracking the exact theoretical curve. This targeted validation confirms the model's accuracy in capturing the specific entanglement dynamics dictated by state parameters, reinforcing its reliability beyond statistical averages.  Even in this challenging high-dimensional scenario $J=5$ (Figure \ref{f33} f), the ensemble model's predictions for the tunable pure state $|\zeta(\theta)\rangle$ maintain excellent agreement with the exact theoretical curve. The close congruence between the predicted (dashed red) and exact (solid blue) lines demonstrates that the model has successfully learned to map the input state features to the correct entanglement value, even as the state traverses a specific path in the Hilbert space.

Collectively, the results presented in Figure \ref{f33} underscore the ensemble model's robust and accurate performance in predicting the negativity of pure quantum states across diverse spin dimensions. While predictive precision exhibits a modest, anticipated decrease with increasing dimensionality for random states due to the expanding complexity of the state space, the model consistently maintains high $R^2$ values and negligible systematic bias. Crucially, its proven ability to accurately reproduce the entanglement characteristics of specific, parametrically defined pure states strongly validates its efficacy and its potential to serve as a reliable tool for characterizing entanglement in pure quantum systems.

\subsection{Performance Evaluation for Mixed Werner States}

\begin{figure}[h!]
	\centering
	\includegraphics[width=0.32\textwidth]{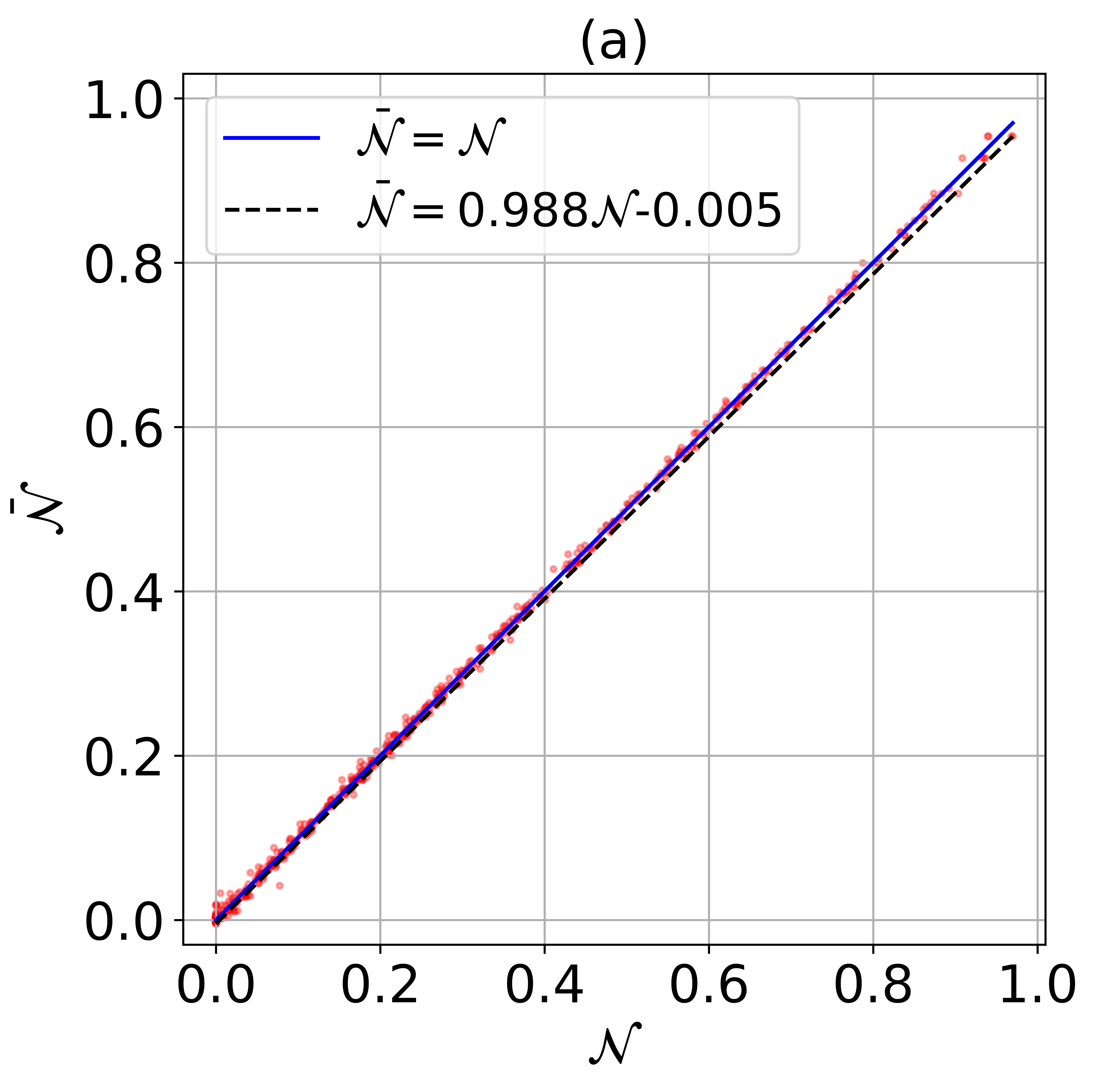}
	\includegraphics[width=0.32\textwidth]{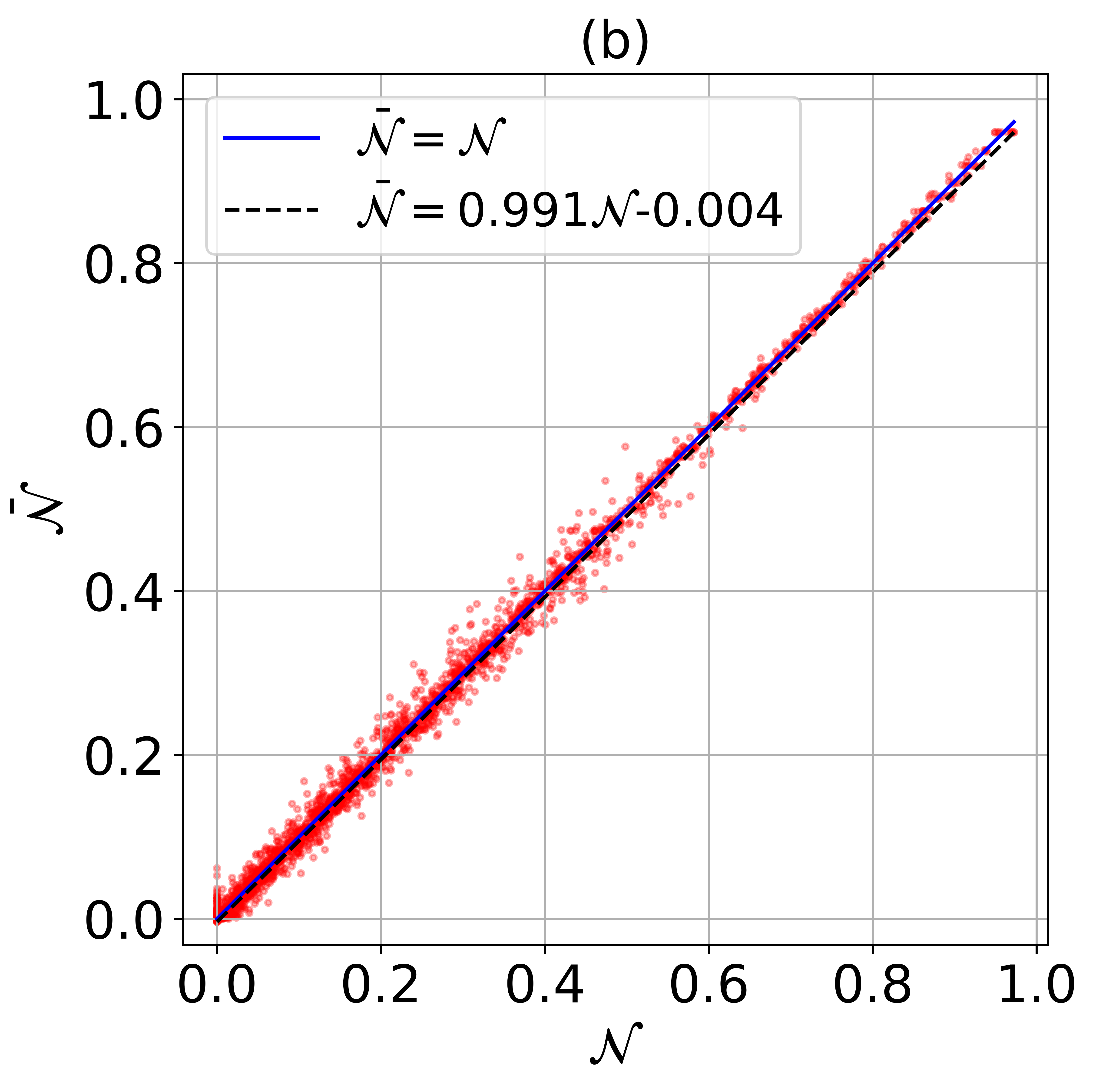}
	\includegraphics[width=0.32\textwidth]{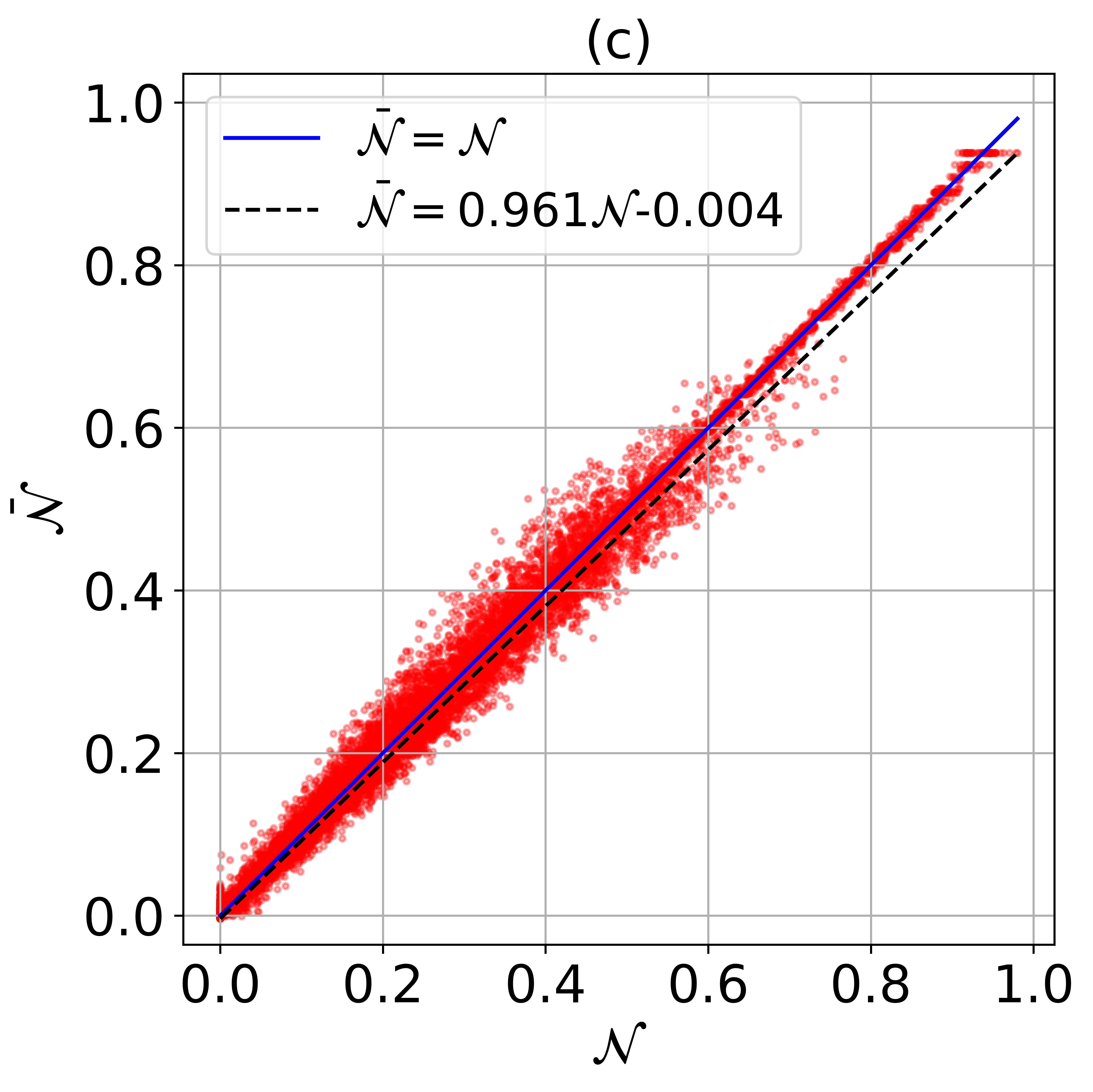} \\
	\includegraphics[width=0.32\textwidth]{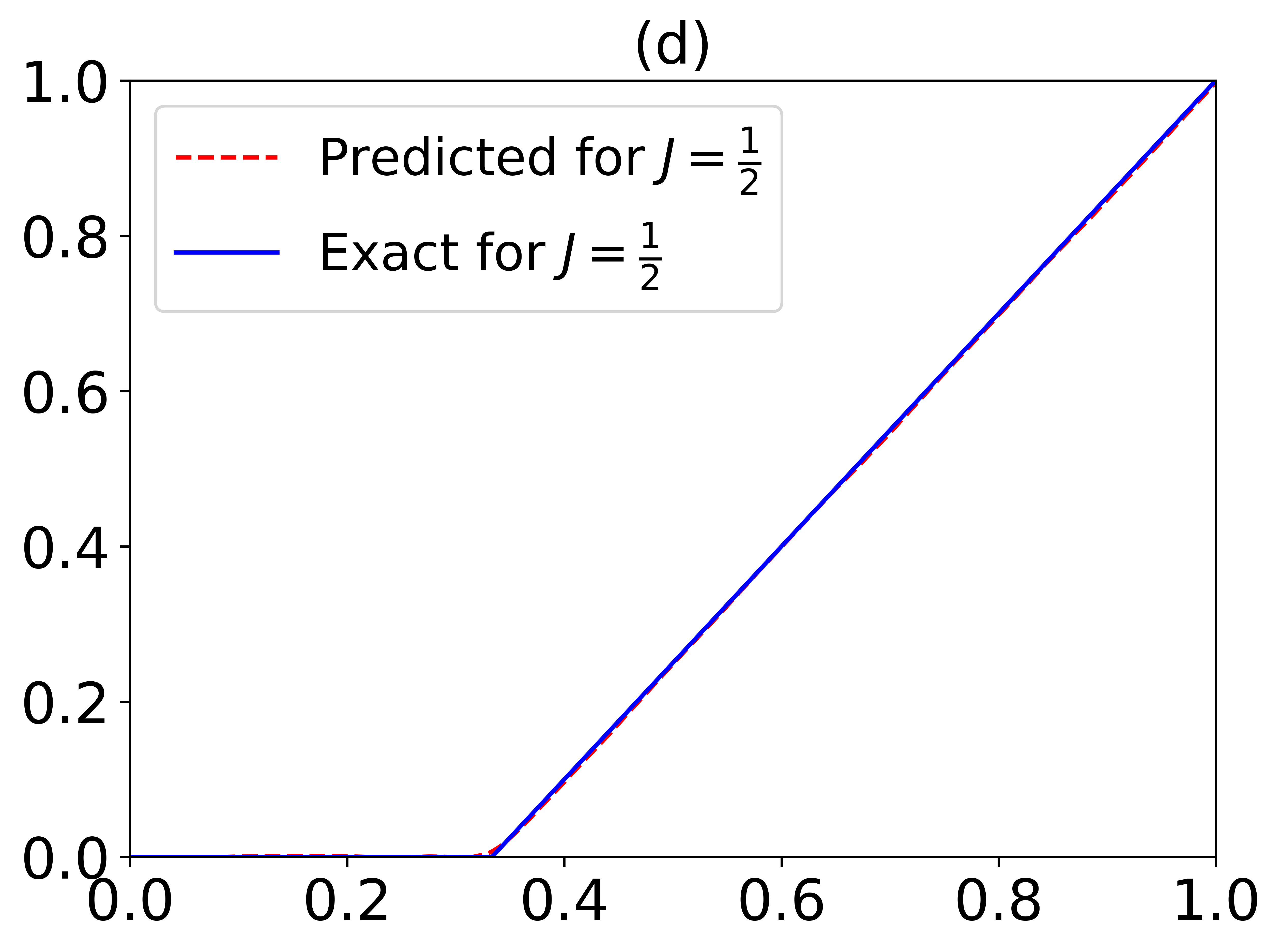}
	\includegraphics[width=0.32\textwidth]{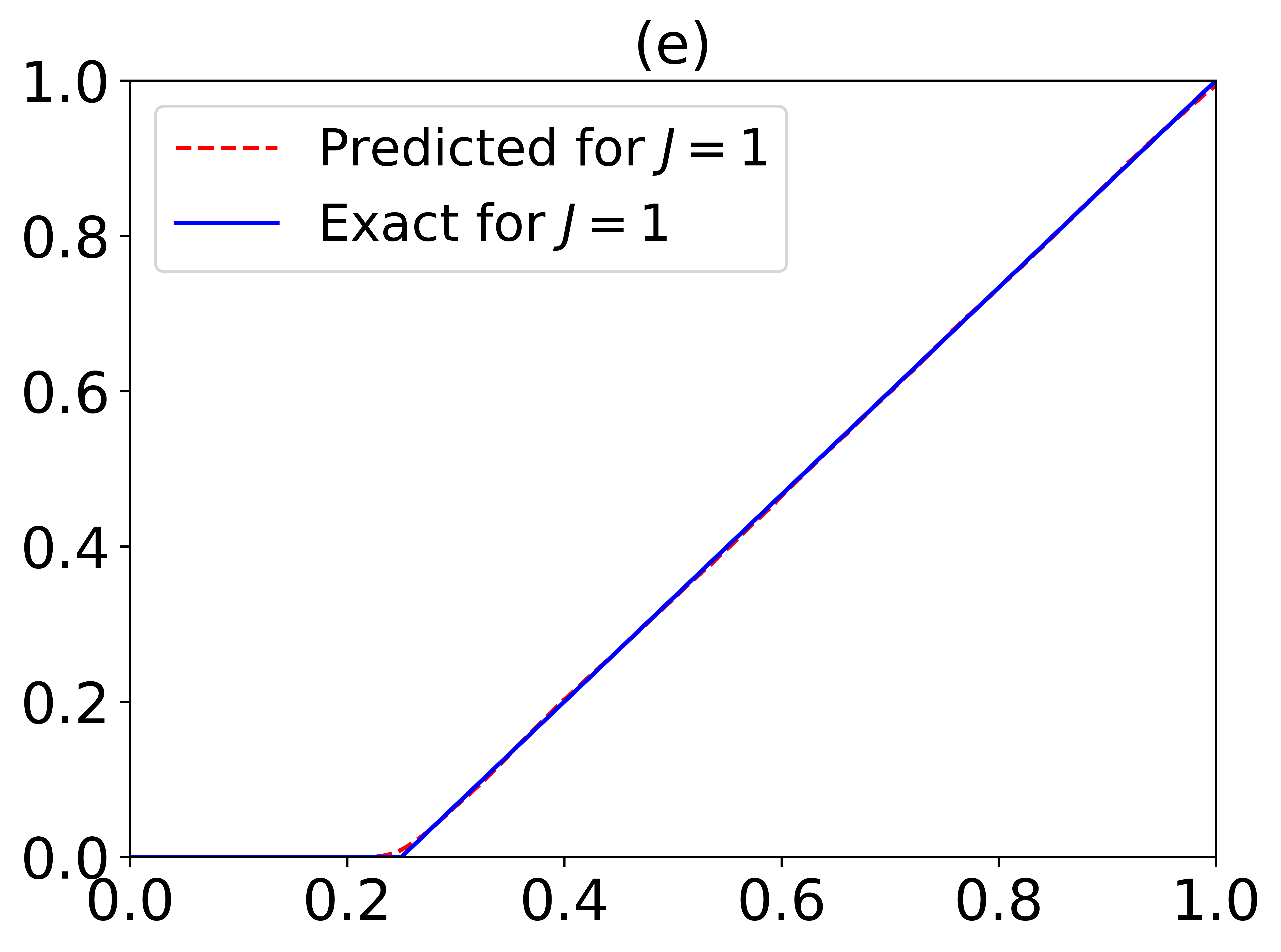}
	\includegraphics[width=0.32\textwidth]{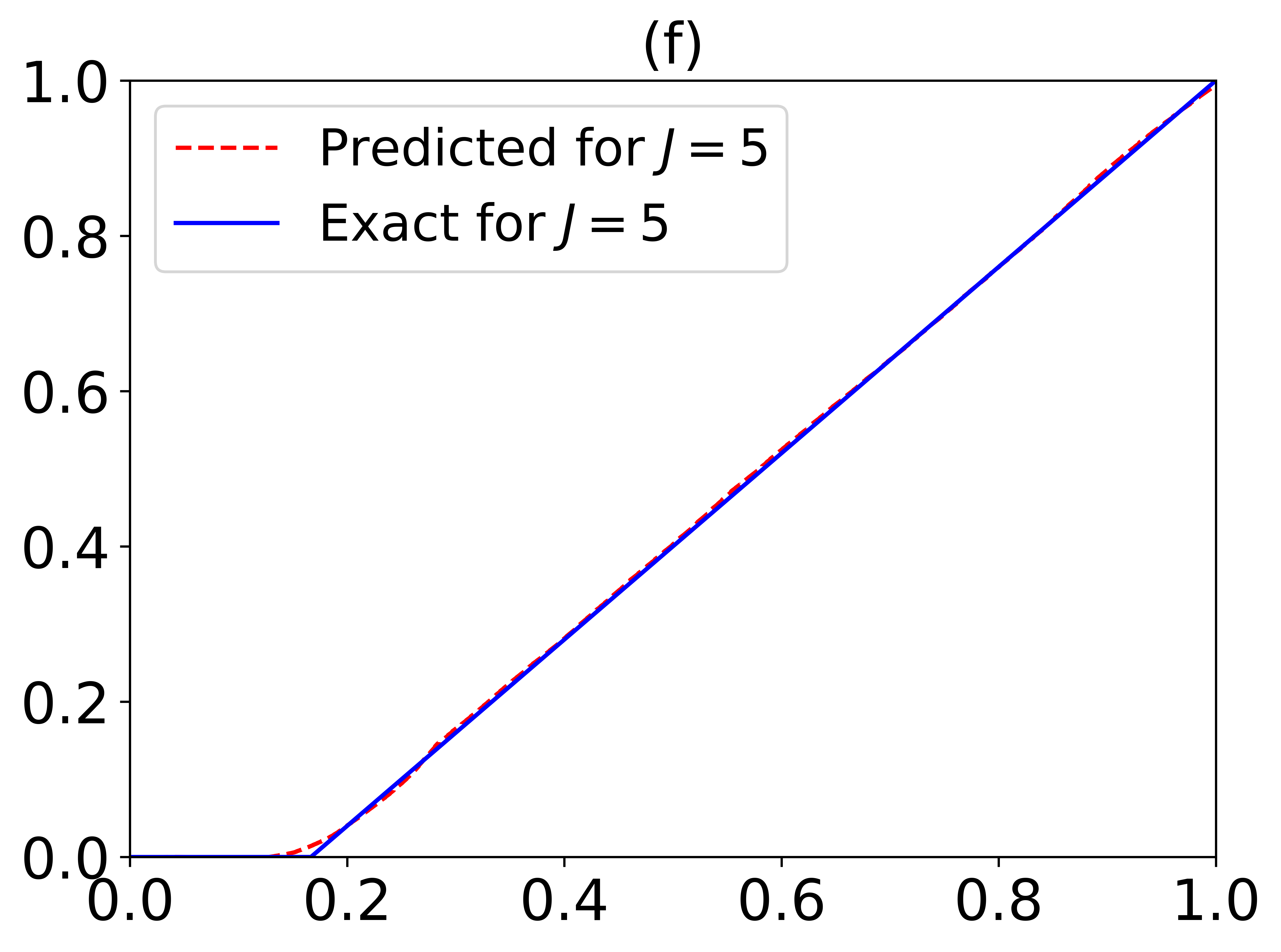}
\caption{Upper plates: Comparison of actual and predicted values for Werner state  (a) \( J = \frac{1}{2} \), (b) \( J = 1 \), (c) \( J =5\). Lower plates: Testing the ensemble model for Werner state  (d) \( J = \frac{1}{2} \), (e) \( J = 1 \), (f) \( J =5\).  }
	\label{f44}
\end{figure}

This subsection assesses the ensemble machine learning model's efficacy in predicting negativity for Werner states, a canonical class of mixed quantum states crucial for benchmarking performance in more realistic scenarios. Werner states that, as defined by Equation 10 represents a convex combination of a maximally entangled pure state and a maximally mixed (white noise) state, controlled by the mixing parameter $\alpha$. The inherent challenge in analyzing such states lies in discerning the quantum correlations responsible for entanglement a midst the classical statistical uncertainty introduced by the mixing process. For the machine learning model to learn from these mixed states, the input features are derived directly from their density matrices. Specifically, for each Werner state, the real and imaginary parts of all elements of its $(2J+1)^2 \times (2J+1)^2$ density matrix $\rho_W$ are flattened into a single feature vector. This provides a complete, unstructured representation of the state from which the model must learn to extract the relevant patterns for predicting negativity.

Figure \ref{f44} (upper panels) illustrates the direct comparison between the actual negativity of randomly generated Werner states (achieved by sampling $\alpha$ values and constructing the corresponding density matrices) and the negativity values predicted by the ensemble model. For $J=0.5$ (Werner states in qubit-qubit systems, Figure \ref{f44} (a)), the model demonstrates commendable predictive power despite the introduction of mixedness. The scatter plot of predicted ($\tilde{\mathcal{N}}$) versus true ($\mathcal{N}$) negativity reveals that data points predominantly align with the ideal $\tilde{\mathcal{N}} = \mathcal{N}$ diagonal. An excellent coefficient of determination ($R^2$) of $0.9997$ is achieved. The linear regression fit, $\tilde{\mathcal{N}} \approx 0.988\mathcal{N} - 0.005$, indicates a robust linear relationship with only minor deviations, underscoring the model's ability to effectively filter noise and extract entanglement-relevant information. The ensemble model maintains its strong performance for qutrit Werner states  ($J=1$, Figure \ref{f44} b), yielding an $R^2$ value of $0.9977$. The linear regression fit, $\tilde{\mathcal{N}} \approx 0.991\mathcal{N} - 0.004$, exhibits minimal deviation from the ideal. This sustained accuracy, even as the Hilbert space dimension increases alongside mixedness, highlights the model's robust learning capabilities. The simultaneous challenges of high dimensionality and statistical mixture make this the most stringent test ($J=5$ Figure \ref{f44}c). Remarkably, the model achieves an outstanding $R^2$ of $0.9928$. The linear regression fit, $\tilde{\mathcal{N}} \approx 0.961\mathcal{N} - 0.004$, remains exceptionally close to the ideal line, signifying very low systematic bias. This high level of accuracy for high-spin mixed states is particularly noteworthy, demonstrating the model's capacity to navigate the combined complexities effectively and validating its scalability.

Further critical validation of the model's understanding of Werner state entanglement is provided by its ability to reproduce the known functional dependence of negativity on the mixing parameter $\alpha$. For $J=0.5$ (Figure \ref{f44} d), the model's predictions (dashed red line) precisely trace the exact theoretical behavior of negativity for Werner states (solid blue line). Crucially, it accurately identifies the critical threshold value of $\alpha$ below which the state becomes separable (negativity vanishes) and correctly captures the linear scaling of negativity above this threshold. Similar outstanding agreement is observed for qutrit Werner states ($J=1$ (Figure \ref{f44} e)). The predicted negativity accurately mirrors the characteristic threshold and subsequent linear increase. The near-perfect congruence between predicted and exact values confirms the model's capability to reliably discern and quantify entanglement across the entire spectrum of mixedness for these systems. Even for these high-spin Werner states, $J=5$ (Figure \ref{f44}f), the model exhibits exceptional fidelity in its predictions across the range of $\alpha$. It accurately reproduces the entanglement threshold and the subsequent linear scaling. This successful benchmark testing, particularly the capture of the qualitative change at the separability threshold, significantly bolsters confidence in the model's predictive power for complex, mixed quantum systems.

In summary, the results presented in Figure \ref{f44} compellingly demonstrate the ensemble model's robust capability to accurately predict entanglement in mixed Werner states across diverse spin dimensions. The consistently high $R^2$ values and close agreement in linear regression fits, even when faced with the dual challenge of high dimensionality and mixedness (as in the $J=5$ case), underscore its effectiveness. The accurate reproduction of the known functional dependence of negativity on the mixing parameter $\alpha$, including the critical separability threshold, further validates that the model has learned essential physical characteristics of entanglement in these systems rather than superficial correlations. This robust performance against mixedness is pivotal, as real-world quantum systems are invariably open and interact with their environment, leading to mixed states. The model's success with Werner states, therefore, significantly enhances its potential for characterizing entanglement in practical quantum information processing and other experimental scenarios.

\section{Conclusion} \label{sec:conclusion}
This study has systematically investigated the application of ensemble machine learning models for the detection and quantification of entanglement, as measured by negativity, in high-spin quantum systems. We successfully developed and evaluated an ensemble stacking with meta-learner combining the strengths of NNs, XGB, and ET, predicting via CB model. The model was rigorously tested on datasets comprising both pure quantum states and mixed Werner states across varying spin dimensions.

Our findings demonstrate that the ensemble machine learning approach provides a highly effective and scalable solution to the computationally intensive task of entanglement estimation. The model consistently achieved high predictive accuracy, as evidenced by low error metrics MSE, MAE, and high coefficients of determination, for both pure and mixed states, even in high-dimensional $J=5$ systems. The performance scalability with dataset size was analyzed, leading to the derivation of an empirical formula that offers a heuristic for estimating data requirements based on system dimensionality and desired predictive accuracy.

Crucially, while individual component models like NNs displayed strong performance based on aggregate metrics, the ensemble model exhibited superior predictive consistency. Visual analysis of prediction scatter plots revealed that the ensemble's predictions were more tightly clustered around the true negativity values with significantly less deviation compared to standalone models. This enhanced reliability, attributed to the inherent error cancellation and variance reduction properties of ensemble learning, underscores the practical advantage of our approach. The ensemble model is not only accurate on average but also provides more trustworthy individual predictions, which is paramount for scientific applications.

Future work could explore the application of this ensemble methodology to other entanglement measures or different classes of quantum states, such as those arising in specific many-body Hamiltonians. Further investigation into optimizing the ensemble architecture, feature engineering for quantum state representation, and incorporating methods for uncertainty quantification within the ensemble predictions could also yield valuable advancements. Ultimately, this research contributes to the growing synergy between machine learning and quantum physics, promising new avenues for discovery in understanding and harnessing the intricacies of the quantum world.

\section*{Appendix}

\appendix
\section{Comparative Performance of Individual Machine Learning Models}
\label{sec:appendix_model_comparison}

This appendix provides a detailed quantitative comparison of the individual machine learning regressors against the proposed stacking ensemble model, justifying its selection as the optimal architecture. The performance metrics for pure states and mixed Werner states are presented in Tables (\ref{tab:appendix_pure_states_final}) and (\ref{tab:appendix_werner_states_final}), respectively.

The data unequivocally demonstrates the superior performance of the ensemble model. While a standalone NN performs well, the ensemble architecture consistently achieves lower error and a higher coefficient of determination across nearly all tested scenarios. This advantage becomes more pronounced in higher-dimensional systems where the prediction task is more challenging.

\begin{table}[htbp]
	\centering
	\renewcommand{\arraystretch}{1}
	\small
	\resizebox{0.8\textwidth}{!}{%
		\begin{tabular}{|l|l|c|c|c|}
			\hline
			\textbf{Model} & \textbf{Performance} & \textbf{J=0.5} & \textbf{J=1} & \textbf{J=5} \\
			\hline
			\multirow{3}{*}{Neural Network} 
			& MSE & 0.0000 & 0.0005 & 0.0014 \\
			& MAE & 0.0033 & 0.0150 & 0.0280 \\
			& $R^2$ & 0.9998 & 0.9916 & 0.9596 \\
			\hline
			\multirow{3}{*}{XGB} 
			& MSE & 0.0003 & 0.0021 & 0.0024 \\
			& MAE & 0.0086 & 0.0309 & 0.0371 \\
			& $R^2$ & 0.9971 & 0.9621 & 0.9294 \\
			\hline
			\multirow{3}{*}{ET} 
			& MSE & 0.0001 & 0.0025 & 0.0066 \\
			& MAE & 0.0046 & 0.0336 & 0.0636 \\
			& $R^2$ & 0.9989 & 0.9544 & 0.8029 \\
			\hline
			\multirow{3}{*}{Random Forest} 
			& MSE & 0.0005 & 0.0033 & 0.0050 \\
			& MAE & 0.0077 & 0.0385 & 0.0530 \\
			& $R^2$ & 0.9970 & 0.9391 & 0.8497 \\
			\hline
			\multirow{3}{*}{Gradient Boosting} 
			& MSE & 0.0061 & 0.0129 & 0.0109 \\
			& MAE & 0.0507 & 0.0872 & 0.0832 \\
			& $R^2$ & 0.9436 & 0.7661 & 0.6715 \\
			\hline
			\multirow{3}{*}{Bagging Regressor} 
			& MSE & 0.0002 & 0.0034 & 0.0050 \\
			& MAE & 0.0077 & 0.0385 & 0.0530 \\
			& $R^2$ & 0.9976 & 0.9391 & 0.8497 \\
			\hline
			\multirow{3}{*}{Ensemble} 
			& MSE & 0.0000 & 0.0002 & 0.0011 \\
			& MAE & 0.0020 & 0.0102 & 0.0245 \\
			& $R^2$ & 0.9999 & 0.9962 & 0.9717 \\
			\hline
		\end{tabular}%
	}
	\caption{Performance comparison of different regression models for various values of $J$ based on MSE, MAE, and $R^2$ score with respect to the pure states.}
	\label{tab:appendix_pure_states_final}
\end{table}

\begin{table}[htbp]
	\centering
	\renewcommand{\arraystretch}{1}
	\small
	\resizebox{0.8\textwidth}{!}{%
		\begin{tabular}{|l|l|c|c|c|}
			\hline
			\textbf{Model} & \textbf{Performance} & \textbf{J=0.5} & \textbf{J=1} & \textbf{J=5} \\
			\hline
			\multirow{3}{*}{Neural Network} 
			& MSE & 0.0000 & 0.0003 & 0.0005 \\
			& MAE & 0.0031 & 0.0103 & 0.0128 \\
			& $R^2$ & 0.9993 & 0.9943 & 0.9902 \\
			\hline
			\multirow{3}{*}{XGB} 
			& MSE & 0.0005 & 0.0009 & 0.0008 \\
			& MAE & 0.0083 & 0.0146 & 0.0156 \\
			& $R^2$ & 0.9872 & 0.9835 & 0.9853 \\
			\hline
			\multirow{3}{*}{ET} 
			& MSE & 0.0002 & 0.0009 & 0.0021 \\
			& MAE & 0.0056 & 0.0169 & 0.0288 \\
			& $R^2$ & 0.9943 & 0.9829 & 0.9592 \\
			\hline
			\multirow{3}{*}{Random Forest} 
			& MSE & 0.0004 & 0.0013 & 0.0018 \\
			& MAE & 0.0083 & 0.0178 & 0.0239 \\
			& $R^2$ & 0.9879 & 0.9768 & 0.9657 \\
			\hline
			\multirow{3}{*}{Gradient Boosting} 
			& MSE & 0.0067 & 0.0057 & 0.0039 \\
			& MAE & 0.0514 & 0.0484 & 0.0400 \\
			& $R^2$ & 0.8150 & 0.8936 & 0.9232 \\
			\hline
			\multirow{3}{*}{Bagging Regressor} 
			& MSE & 0.0004 & 0.0013 & 0.0018 \\
			& MAE & 0.0083 & 0.0178 & 0.0239 \\
			& $R^2$ & 0.9878 & 0.9768 & 0.9658 \\
			\hline
			\multirow{3}{*}{Ensemble} 
			& MSE & 0.0000 & 0.0001 & 0.0003 \\
			& MAE & 0.0013 & 0.0062 & 0.0111 \\
			& $R^2$ & 0.9997 & 0.9977 & 0.9928 \\
			\hline
		\end{tabular}%
	}
	\caption{Performance comparison for mixed Werner states ($J=0.5, 1, 5$).}
	\label{tab:appendix_werner_states_final}
\end{table}

For instance, consider the high-spin pure states with $\mathbf{J=5}$ (Table \ref{tab:appendix_pure_states_final}). The ensemble model achieves an \textbf{MSE of 0.0011} and an $\mathbf{R^2}$ \textbf{of 0.9717}, a clear improvement over the standalone NN's MSE of 0.0014 and $R^2$ of 0.9596. This trend is even more significant for the mixed Werner states (Table \ref{tab:appendix_werner_states_final}), where for $\mathbf{J=5}$, the ensemble's \textbf{MSE of 0.0003} is substantially better than the NN's 0.0005.

This quantitative superiority is visually corroborated by the prediction scatter plots presented in the main text (Figures \ref{f33} and \ref{f44}). The lower MSE and MAE values of the ensemble model directly correspond to the visibly tighter clustering of predictions around the true negativity values. This demonstrates a reduction in both bias and variance. The enhanced performance is attributed to the stacking architecture, where the CatBoost meta-learner effectively combines the predictions of the diverse base models (NN, XGB, ET), canceling out their uncorrelated errors and leveraging their collective strengths.

In conclusion, the selection of the ensemble model is justified not only by the theoretical principles of variance reduction but also by its \textbf{demonstrably superior quantitative performance} in this study. It provides a more accurate, reliable, and robust tool for entanglement quantification, which is paramount for scientific applications.

\section*{Data Availability}
The data and code that support the findings of this study are openly available in a GitHub repository at \url{https://github.com/Amr0MEid/Entanglement-Detection-Using-Ensemble-Learning} and are permanently archived on Zenodo under the DOI \url{https://doi.org/10.5281/zenodo.16010784}.

\bibliographystyle{unsrt}
\bibliography{bm}

\end{document}